\DocumentMetadata{testphase=new-or-1}

\documentclass[twocolumn]{aastex63}

\usepackage[T1]{fontenc}

\usepackage{etoolbox}
\apptocmd{\thebibliography}{\interlinepenalty 10000\relax}{}{}
\graphicspath{{./}{figs/}}

\usepackage{mathtools}
\usepackage{amsmath}
\usepackage{amsfonts}
\usepackage{amssymb}
\usepackage{appendix}
\usepackage{needspace}
\usepackage{xspace}
\usepackage{multirow}
\usepackage{physics}

\newif\ifref
\reftrue
\reffalse
\definecolor{darkred}{rgb}{0.75, 0, 0}
\newcommand{\mb}[1]{\ifref\textcolor{darkred}{#1}\else #1\fi}

\newcommand{\Rsun}{\ensuremath{\,\mathrm{R_\odot}}\xspace} 
\newcommand{\Msun}{\ensuremath{\,\mathrm{M_\odot}}\xspace} 
\newcommand{\Lsun}{\ensuremath{\,\mathrm{L_\odot}}\xspace} 
\newcommand{\MBH}{$M_{\textrm{BH}}$}

\let\oldsection\section
\renewcommand{\section}{\needspace{3\baselineskip}\oldsection}
\let\oldsubsection\subsection
\renewcommand{\subsection}{\needspace{2\baselineskip}\oldsubsection}

\received{\today}
\submitjournal{The Astrophysical Journal}

\shorttitle{Solar Evolution Models with a Central Black Hole}
\shortauthors{Bellinger et al.}

\begin{document}
\title{Solar evolution models with a central black hole}

\correspondingauthor{E.\ P.\ Bellinger}
\email{earl.bellinger@yale.edu}

\author[0000-0003-4456-4863]{Earl P.~Bellinger} 
\affiliation{Max Planck Institute for Astrophysics, Garching, Germany}  
\affil{Department of Astronomy, Yale University, CT, USA} 
\affil{Stellar Astrophysics Centre, Department of Physics and Astronomy, Aarhus University, Denmark} 

\author[0000-0000-0000-0000]{Matt E. Caplan} 
\affiliation{Department of Physics, Illinois State University, IL, USA}

\author[0000-0003-2012-5217]{Taeho Ryu} 
\affiliation{Max Planck Institute for Astrophysics, Garching, Germany}  
\affiliation{Physics and Astronomy Department, Johns Hopkins University, Baltimore, MD, USA}

\author[0000-0001-8835-8733]{Deepika Bollimpalli} 
\affiliation{Max Planck Institute for Astrophysics, Garching, Germany}  

\author[0000-0002-4773-1017]{Warrick H.~Ball} 
\affiliation{School of Physics and Astronomy, University of Birmingham, Birmingham, UK}
\affil{Advanced Research Computing, University of Birmingham, Birmingham, UK}

\author[0000-0002-1528-1920]{Florian K\"uhnel}
\affiliation{Max Planck Institute for Physics, Munich, Germany}  

\author[0000-0003-3441-7624]{R. Farmer} 
\affiliation{Max Planck Institute for Astrophysics, Garching, Germany}  

\author[0000-0001-9336-2825]{S.~E.~de~Mink}
\affiliation{Max Planck Institute for Astrophysics, Garching, Germany}  
\affiliation{Anton Pannekoek Institute for Astronomy and GRAPPA, University of Amsterdam, The Netherlands}

\author[0000-0001-5137-0966]{J{\o}rgen Christensen-Dalsgaard} 
\affil{Stellar Astrophysics Centre, Department of Physics and Astronomy, Aarhus University, Denmark} 

\begin{abstract}
\cite{1971MNRAS.152...75H} proposed that the Sun may harbor a primordial black hole whose accretion supplies some of the solar luminosity. 
Such an object would have formed within the first $1\,$s after the Big Bang with the mass of a moon or an asteroid. 
These light black holes are a candidate solution to the dark matter problem, and could grow to become stellar-mass black holes (BHs) if captured by stars. 
Here we compute the evolution of stars having such a BH at their center. 
We find that such objects can be surprisingly long-lived, with the lightest black holes having no influence over stellar evolution, while more massive ones consume the star over time to produce a range of observable consequences. 
Models of the Sun born about a BH whose mass has since grown to approximately $10^{-6}$~\Msun{} are compatible with current observations.
In this scenario, the Sun would first dim to half its current luminosity over a span of 100~Myr as the accretion starts to generate enough energy to quench nuclear reactions. 
The Sun would then expand into a fully-convective star, where it would shine luminously for potentially several Gyr with an enriched surface helium abundance, first as a sub-subgiant star, and later as a red straggler, before becoming a sub-solar-mass BH. 
We also present results for a range of stellar masses and metallicities. 
The unique internal structures of stars harboring BHs may make it possible for asteroseismology to discover them, should they exist. 
We conclude with a list of open problems and predictions. 
\end{abstract}

\keywords{stellar evolution --- primordial black holes --- dark matter} 

\section{Introduction} \label{sec:intro} 
The dark matter problem has now become serious \citep{1975ApJ...201..489C}. 
Numerous lines of evidence---such as from galaxy rotation curves \citep{1933AcHPh...6..110Z, 1970ApJ...159..379R}, the large-scale structure of the universe \citep{1996ApJ...457L..51H}, and the cosmic microwave background \citep[][]{2016A&A...594A..13P}---indicate that most of the matter in the Universe is invisible. 
Yet despite nearly a century of research, the origin of this matter remains unknown, and no compelling evidence has emerged for a solution. 
Leading candidates include novel particles such as axion-like particles, weakly interactive massive particles (WIMPs), and sterile neutrinos \citep[for reviews, see e.g.][]{2010ARA&A..48..495F,2018PhRvD..98c0001T}. 

One proposed solution is compact objects formed within the first second after the Big Bang: primordial black holes (PBHs, \citealt{1974MNRAS.168..399C}; see \citealt{2021arXiv211002821C, 2022arXiv221105767E} for reviews). 
These can arise from inhomogeneities in the initial state of the Universe and result in black holes (BHs) whose masses are proportional to the cube of the time of their formation \citep{1975ApJ...201....1C}. 
PBHs are considered attractive because unlike many other solutions, they require no modification to the standard model of particle physics. 
\citet{Carr2023Review} enumerate several hints that may point to the probable production of at least some PBHs during the aftermath of the Big Bang, but it is currently unclear how many were produced, and whether they are sufficient in number to explain the dark matter. 
Ultimately, dark matter may be a combination of various solutions, possibly including PBHs~\citep[e.g.,][]{2016AstL...42..347E, 2021MNRAS.506.3648C}. 

There are several promising avenues for detecting PBHs, though disentangling them from more mundane astrophysical signals is a challenge. 
Gravitational-wave observations from LIGO/Virgo present one such opportunity \citep{Clesse:2016vqa, 2020JCAP...09..022J, Escriva:2022bwe, Carr2023Review}. 
Four candidate binary BH mergers may have a subsolar component \citep{2021arXiv210511449P}, which, if confirmed, could not possibly come from stars, and would be a clear indication of a primordial origin. 
There are also several with progenitor masses between 60 and 110~\Msun{} \citep{2021arXiv211103606T}, which are within the pair-instability mass gap \citep{2019ApJ...887...53F, 2022ApJ...937..112F} and hence cannot originate directly from the collapsing cores of stars. 
Plausible explanations include repeated mergers \citep{2019PhRvD.100d3027R, 2022ApJ...927..231F} or growth from smaller seeds in a nuclear star cluster \citep{2021MNRAS.501.1413N}, but these may also be PBHs. 

Another possible route to the discovery of PBHs may be through gravitational microlensing. 
The OGLE survey has detected numerous compact bodies of very low mass ($10^{-4} - 10^{-10}$~\Msun{}) that may be free-floating planets or PBHs \citep{2019PhRvD..99h3503N}. 
The PBH scenario also leads to a range of cosmological and dynamical consequences that may lend support for their production, such as cosmological background correlations and their effects on the reionisation history of the Universe \citep{2013ApJ...769...68C, 2005Natur.438...45K, 2018RvMP...90b5006K, Kashlinsky:2016sdv, 2020JCAP...07..022H, Boldrini:2019isx, 2022ApJ...926..205C, Carr2023Review}.

The solar system also provides some tools for detecting PBHs. 
The Earth, Moon, and Sun can be used as transient PBH detectors, as typically-assumed velocities would have PBHs passing through these objects faster than escape velocity, which would lead to unique dynamics \citep{2022arXiv220912415L}, oscillations \citep{2011PhRvL.107k1101K, 2012ApJ...751...16L}, and craters \citep{2021MNRAS.505L.115Y}. 
\citet{2020PhRvL.125e1103S} have argued that the hypothesized Planet~9 \citep{2016AJ....151...22B} could be a PBH, which could be confirmed by annihilation signals. 

The Milky Way is expected to contain $\sim$100 million BHs from normal stellar evolution pathways, with the average BH being at a distance of 21~pc ($\sim 10^6$~au) \citep{2022MNRAS.516.4971S}. 
If PBHs at the classical Hawking evaporation limit ($\sim 10^{-18}$~\Msun{}) constitute the dark matter, this number increases to $\sim 10^{30}$ BHs at an average distance of $\sim$1~AU. 
This would be far more numerous and far more densely spaced than stars, raising the possibility of their capture by stars \citep{2021PhRvD.104l3031I, 2022ApJ...932...46I} or star-forming clouds. 

If PBHs comprise all or part of the galactic dark matter halo, then they may have orbits with random inclinations and velocities from a Maxwell-Boltzmann distribution. 
PBHs in the slow tail of this distribution, if aligned with the orbits of stars or star-forming regions, have some chance of being captured. 
\citet{2019JCAP...08..031M} have found that the probabilities of capture by a star are low, as a PBH on average falls in faster than the stellar escape velocity and hence transits the star with only negligible dissipation from dynamical friction and accretion. 
The capture of PBHs during star formation is much more likely due to the time-dependent gravitational potential of the adiabatically collapsing cloud \citep{2013PhRvD..87b3507C,2014PhRvD..90h3507C, 2023NewA..10302057E}. 
PBH capture in a star-forming region is even more likely in dwarf galaxies due to their lower mean velocities as well as in the early universe, which has formed the basis for additional observational constraints on PBH dark matter \citep{esser2022constraints, 2022MNRAS.517...28O}. 

A low-velocity PBH captured by a star would slowly consume it \citep{1975ApJ...201..489C, 1981A&A....99...31P, 2022MNRAS.517...28O}. 
At early times, the PBH may accrete at a Bondi-like rate ($\textrm{d}M_{\textrm{BH}}/\textrm{d}t \propto M_{\textrm{BH}}^2$, with $M_{\textrm{BH}}$ the mass of the BH), where accretion is limited by the sound speed in the center of the star. 
It may later transition to an Eddington-like accretion (${\textrm{d}M_{\textrm{BH}}/\textrm{d}t \propto M_{\textrm{BH}}}$) if the radiation pressure can stall the freefall accretion, but this is uncertain. 

Following Hawking's suggestion that the Sun may harbor a PBH \citep{1971MNRAS.152...75H}, and inspired by the then-unresolved solar neutrino problem \citep{1973ApL....13...45S}, \citet{1975ApJ...201..489C} created the first solar BH models from partial evolutionary sequences. 
\citet{1981MNRAS.194..475T} and \citet{1982MNRAS.199..833F, 1984MNRAS.206..589F} considered optically-thick, Bondi-type accretion and the role of gas pressure, and \citet{1995MNRAS.277...25M, 1995MNRAS.277...11M} analyzed the effects of convection, rotation, and magnetic fields. 
On the other end of the mass spectrum, BHs inside of supermassive stars ($M>1000$~\Msun{}), \emph{quasi-stars}, have been evoked to explain the origin of supermassive BHs in the early Universe \citep{2008MNRAS.387.1649B, 2010MNRAS.402..673B, 2010MNRAS.409.1022V, 2011MNRAS.414.2751B, 2012MNRAS.421.2713B}. 
Several authors have studied the growth of PBHs inside neutron stars \citep{2014PhRvD..90d3512K, 2021PhRvD.103h1303B, 2021PhRvD.104l3021S, 2021PhRvD.103j4009R}. 

In this work, we carry these models forward and consider the evolution of stars having $0.8 - 100$~M$_\odot$ that have formed about a BH spanning all possible masses up to the stellar mass, including the first full numerical evolutionary simulations of solar-like stars with a central BH. 
We utilize a 1D stellar evolution code that makes use of modern opacities and other microphysics in the calculations, as well as multiple treatments of the accretion. 
We describe the physics and implementation in Section~\ref{sec:methods}. 
We present detailed numerical results on solar models in Section~\ref{sec:solar}, and analytically explore a grid of stellar models in Section~\ref{sec:stellar}. 
We discuss limitations of our model in Section~\ref{sec:future}, and list several open problems. 
Finally, we conclude the paper in Section~\ref{sec:conclusions}.

\section{Methods} \label{sec:methods}

In order to simulate the evolution of a star under the hypothesis that it has a central black hole, we treat the BH as a point mass at the core and solve the 1D stellar structure equations with modified inner boundary conditions \citep{1975ApJ...201..489C, 2012PhDT.........1B, 2023arXiv230507337F} through an extension to the \textsc{Mesa} \citep[Modules for Experiments in Stellar Astrophysics, r23.05.1;][]{2011ApJS..192....3P, 2013ApJS..208....4P, 2015ApJS..220...15P, 2018ApJS..234...34P, 2019ApJS..243...10P, 2022arXiv220803651J} stellar evolution code. 

In the standard equations of stellar structure, the mass $M$, radius $R$, and luminosity $L$ are all zero at the centerpoint. 
We set the mass at the inner boundary to \MBH{} plus the mass of a cavity around the BH from which it accretes. 
This models the Bondi sphere, where the infall velocity is greater than the speed of sound in the medium \citep{1952MNRAS.112..195B}. 
We take the Bondi radius as the radial inner boundary condition, defined as 
\begin{equation}
    R_{\rm B} 
    = 
    2\, 
    \frac{G M_{\textrm{BH}}}{c_s^{2}}
    \, ,
\end{equation}
where $c_s$ is the adiabatic sound speed at $R_B$. 
We take the mass in this region as $(8\pi/3) \rho R_{\textrm{B}}^3$, assuming a density profile of $\rho \propto r^{-3/2}$ \citep{2012PhDT.........1B}.

The luminosity generated from accretion is given by
\begin{equation}
    L 
    = 
    \dfrac{\epsilon}{1-\epsilon}\, 
    \frac{\textrm{d}M_{\textrm{BH}}}{\textrm{d}t}\, 
    c^2
    \, ,
\end{equation}
where $c$ is the speed of light and $\epsilon$ is the radiative efficiency. 
As a first case, we study a fixed radiative efficiency at the canonical ${\epsilon = 0.08}$ roughly corresponding to reaching the innermost stable orbit around a Schwarzschild BH, but this efficiency is highly uncertain (see Section~\ref{sec:super}). 
For comparison, the efficiency of nuclear fusion is an order of magnitude smaller at $\sim 0.007$, and the efficiency of a rotating Kerr BH can be much larger. 

We take the luminosity as the lesser of the Eddington luminosity $L_{\rm E}$ and a Bondi-like convective luminosity $L_{\rm B}$ \citep{2012PhDT.........1B}, defined as
\begin{align}
    L_{\rm E} 
    &= \label{eq:edd} 
    4\pi\, 
    \dfrac{c}{\kappa}\, 
    G M_{\textrm{BH}}
    \, ,
    \\
    L_{\rm B} 
    &= \label{eq:bondi} 
    16\pi \eta\, 
    \dfrac{\rho}{c_s \Gamma_1}\, 
    \left(G M_{\textrm{BH}} \right)^2 
    \, ,
\end{align}
where $\kappa$ is the opacity and $\Gamma_1$ is the first adiabatic exponent. 
The efficiency of convection is denoted $\eta$ and we take it to be $0.1$ \citep{2012PhDT.........1B}. 
We also consider a case without Eddington-limited accretion, in which the luminosity is given only by Equation~\ref{eq:bondi}. 
We adopt the widely-used \textsc{Opal} \citep[Opacity Project at Livermore,][]{1996ApJ...464..943I, 2002ApJ...576.1064R} equation of state and opacities. 
To give typical values, standard models of the present Sun have ${\rho = 150~\textrm{g}~\textrm{cm}^{-3}}$, ${c_s = 5\times 10^7~\textrm{cm}~\textrm{s}^{-1}}$, ${\kappa = 1.5~\textrm{cm}^2~\textrm{g}^{-1}}$, and ${\Gamma_1 \simeq 5/3}$. 
Note that the metal-rich plasma of the solar core is much more opaque than the commonly adopted electron-scattering opacity for ionized hydrogen ($\kappa = 0.4~\textrm{cm}^2~\textrm{g}^{-1}$). Values for zero-age main sequence (ZAMS) stars across our chosen mass range are given in Appendix~\ref{app:zams}. 

The mass of the system decreases over time by the equivalent of the energy converted into radiation. 
We limit the timestep to $M_{\textrm{BH}} / (\textrm{d}M_{\textrm{BH}}/\textrm{d}t)$ in order to stably resolve the accretion.  
We include a small amount of exponentially-decaying convective overshoot in order to smooth convective boundaries. 
We do not consider element diffusion or mass loss by stellar winds. 
We also do not model the late stages of the consumption of the star. 
Our implementation and models are publicly available\footnote{\url{https://github.com/earlbellinger/black-hole-sun}} \mb{\citep{bhsun-zenodo}}.

\section{Results}
\subsection{Solar models} \label{sec:solar}%

The Sun is formidable and destroying it is generally considered to be a difficult problem. 
Stellar evolution models without a central BH suggest that the Sun will evolve into a black dwarf and live on essentially forever, possibly only destroyed by proton decay in the far distant future \citep{2007S&T...113f..32L, 2020MNRAS.497.4357C}.
As a fiducial first case, we characterize the evolution of the Sun with a central PBH until its premature demise. 

We calibrated solar models with a central black hole with masses above the classical Hawking evaporation limit (\MBH{}~$>10^{-20}$~\Msun{}). 
These models have the observed present-day properties of the Sun, but get some fraction of their luminosity from accretion onto a black hole. 
For each PBH mass, we adjusted the initial helium abundance and mixing-length parameter until the models reached $1$~\Rsun{} and $1$~\Lsun{} with $\textrm{[Fe/H]}=0$ within a tolerance of $10^{-7}$ at the solar age of $4.572$~Gyr. Models with an initial \MBH{}~$\lesssim 10^{-11}$~\Msun{} converged; models with greater mass are either overluminous or consume the Sun before its present age. 
The converged models correspond to PBHs that formed in the first femtoseconds of cosmic time: approximately $10^{-24}$~--~$10^{-14}$~s after the Big Bang, following the inflationary epoch and preceding the quark epoch. 

We find that a PBH spanning a wide range of masses (a current \MBH{}~$\leq 10^{-6}$~\Msun) could exist inside of the present Sun without significant modification to its present structure. 
Because the masses of these BHs are so small and the luminosity from their accretion is much less than the nuclear luminosity, they change the sound speed, density, and nuclear reaction rates in the Sun by much less than a per cent. 
These models are therefore compatible with present-day helioseismic and solar neutrino observations. 
This mass window coincides with the PBH mass range that could still constitute the entirety of the dark matter~\citep{2021arXiv211002821C}. 

\begin{figure*}
	\centering
        \includegraphics[width=\textwidth]{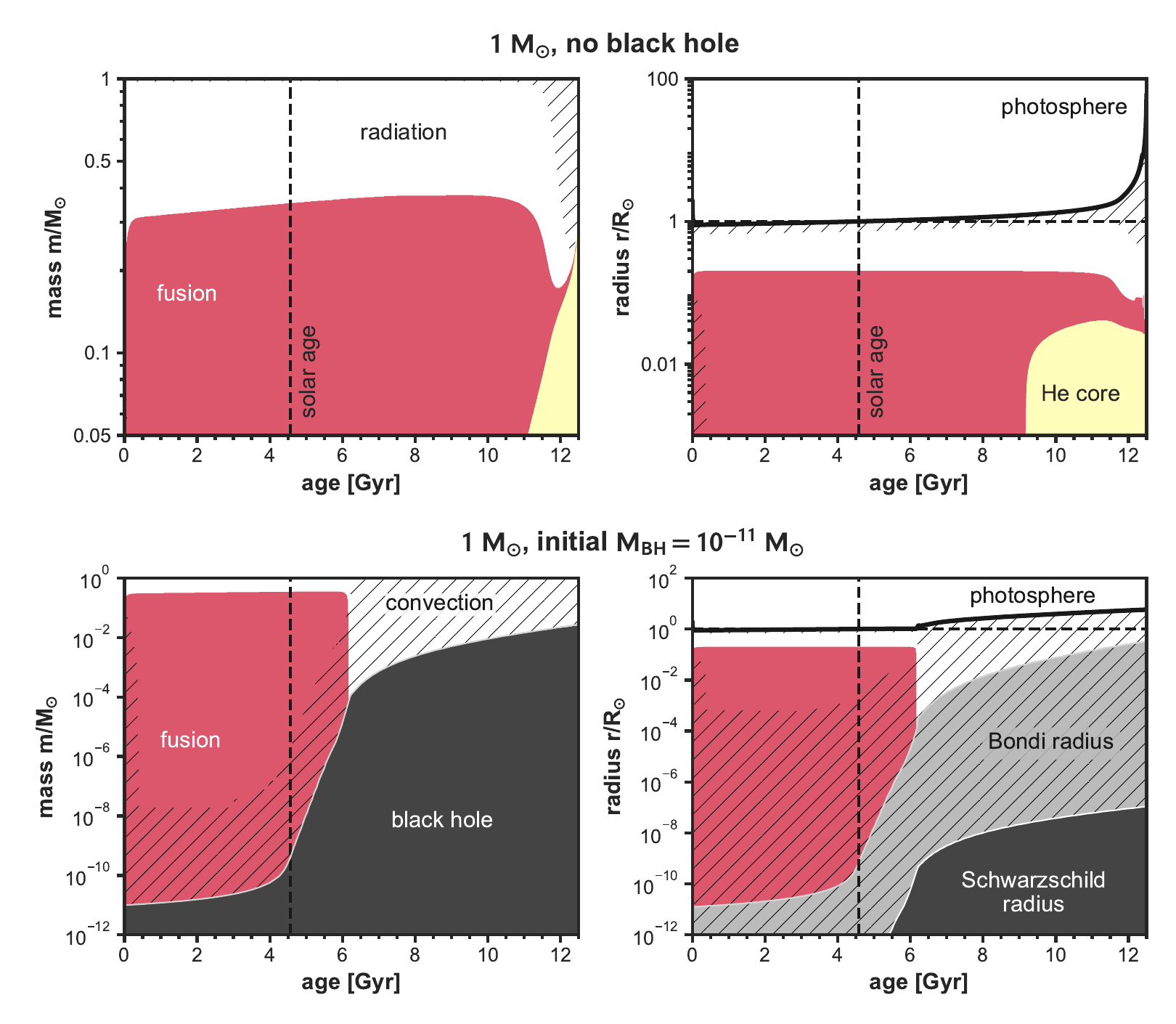}%
	\caption{Kippenhahn diagrams showing the evolution of the interior of the Sun with and without a central black hole. 
    The left panels show the mass distribution, with regions of energy generation and transport indicated. 
    The right panels show the radial distribution, with the radius of the photosphere (black line) and the solar radius (horizontal dashed line) indicated. 
    The top panels correspond to a normal solar evolution model evolved through the main sequence until core hydrogen exhaustion and up through hydrogen shell burning as a red giant. 
    The bottom panels show a model that is consistent with the present Sun with a black hole growing at its center. 
    Nuclear fusion (red) provides the bulk of the solar luminosity until the black hole is of sufficient mass to quench the reactions. 
    The black hole drives convection (hatches), which mixes the innermost parts of the core, and eventually the entire star. 
    Note the differences in y-axis scale between the panels. 
    \label{fig:kippenhahn}
    }
\end{figure*}

\begin{figure*}
    \includegraphics[width=0.33\textwidth]{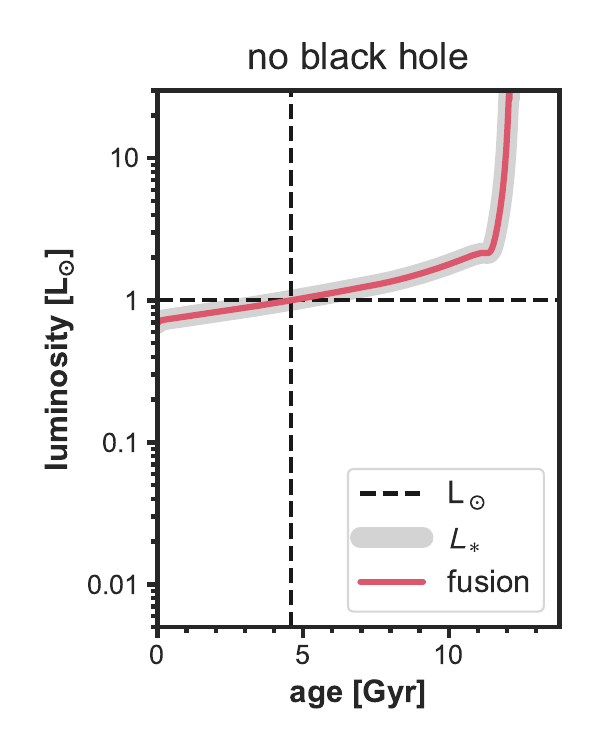}%
    \includegraphics[width=0.33\textwidth]{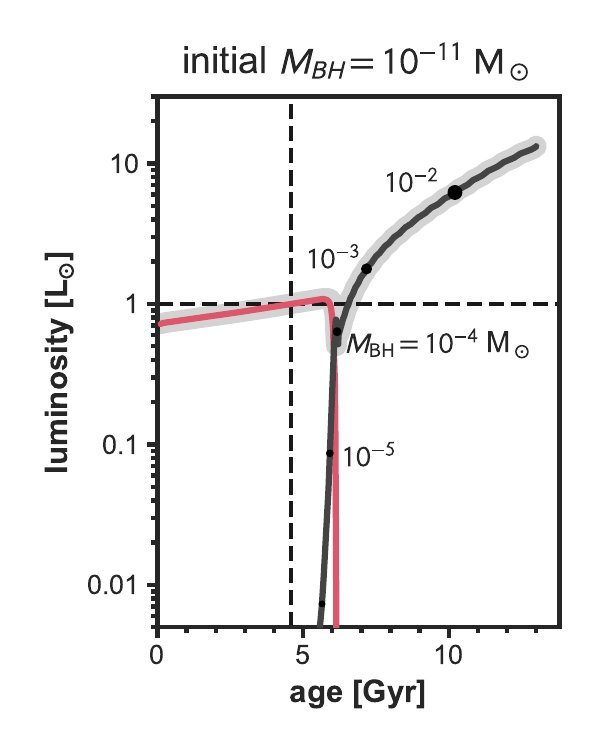}%
    \includegraphics[width=0.33\textwidth]{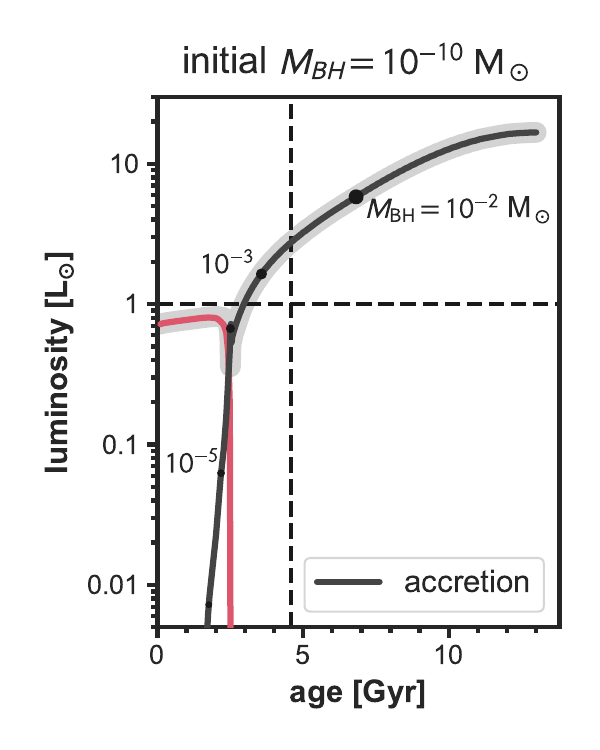}%
    \caption{
        \textsc{Mesa} simulations of the past and future luminosity of the Sun assuming no central black hole (left), a central black hole of initial mass $10^{-11}$~\Msun{} (center), and one of $10^{-10}$~\Msun{} (right). 
        The solar age and luminosity are indicated as well as the mass of the black hole. 
        Unlike the first two models, the model on the right is obviously not compatible with the present Sun. 
        In both of the right panels, it takes $\sim$~1~Gyr for the black hole to transition from supplying 1\% to supplying 100\% of the solar flux. 
        When the black hole approaches the point of supplying 20\% of the solar luminosity, nuclear reactions begin to wane and eventually cease. The Sun then dims to half its current luminosity over the span of 100~Myr. 
        Over the course of the next 300~Myr, the black hole grows to a mass of $5\times10^{-5}$~\Msun{}, at which point it produces the full solar luminosity. Soon thereafter, the Sun expands into a red giant, and does so at a much greater pace than in the normal case: 
        instead of taking 6 Gyr for the Sun to double in luminosity, it is achieved with aid of the black hole in $\sim$300~Myr. 
        Finally, the accretion luminosity drives the Sun to slowly expand over the course of several Gyr until it produces a maximum of more than 10~\Lsun{} before the simulation terminates. 
        The black hole at the bottom of the middle panel having a mass of $10^{-6}$~\Msun{} is depicted with its actual Schwarzschild radius ($\sim$3~mm); more massive black holes are too large to show (e.g., $\sim$30~m at $10^{-2}$~\Msun{}) on this plot. 
    \label{fig:luminosity}}
\end{figure*}

Figure~\ref{fig:kippenhahn} shows the evolution and fate of the Sun if it had formed about a PBH with an initial mass of $10^{-11}$~\Msun{}. 
Over the lifetime of the Sun from the pre-main sequence (PMS) until the present day (4.572~Gyr), the PBH increases its mass tenfold as it accretes the solar plasma. The BH drives convection in the core of the Sun and mixes its innermost regions, but otherwise causes little change to its outward appearance. 
The fate of the Sun, on the other hand, changes dramatically. 
By an age of 7 Gyr, the BH consumes 0.1\% of the solar mass. 
The solar core now cools, causing nuclear reactions to cease. 
Soon thereafter, the luminosity from accretion starts to vastly exceed what was once the luminosity from nuclear fusion (see Figure~\ref{fig:luminosity}). 

Next, the large accretion luminosity causes the star to become fully convective and puff into a red giant. 
The partially fused core gets fully mixed back into the envelope, leading to a star with an extremely high surface helium abundance. 
The expansion of the Sun halts at a maximum of $\sim$0.03~AU, and thus the Earth is saved from being engulfed by the red giant Sun. 
Earth's oceans nevertheless still boil off as its blackbody temperature reaches upwards of 530~K ($\sim 250^\circ$C). 

As acoustic stellar oscillation frequencies are sensitive to the sound speed of the core and hence the core temperature, the initial transition to lower core temperatures would likely bear an observable seismic signature \citep[e.g.,][]{1989nos..book.....U, 2010aste.book.....A, basuchaplin}, thus providing an early warning for the changes that are to come. 
Discovery of solar $g$ modes would also help to falsify this model, as these as yet undetected oscillations would grant insights into the deepest layers of the Sun. 

In this model, the Sun has about 8~Gyr after the present day remaining before it will become a black hole, including a >2~Gyr timespan in which the BH mass is greater than 1\% of the solar mass. 
In the Kippenhahn diagram (Figure~\ref{fig:kippenhahn}), two distinct changes of slope are evident in the BH growth: at 5~Gyr (\MBH{}~$\simeq$~$10^{-10}$~\Msun{}) when the accretion transitions from the Bondi rate to Eddington; and again at 7~Gyr (\MBH{}~$\simeq$~$10^{-4}$~\Msun{}), when the accretion takes over to supply the full luminosity of the star. 

Figure \ref{fig:luminosity} additionally shows the evolution of the Sun if it formed about a PBH with a greater initial mass of $10^{-10}$~\Msun{}. 
Here the situation is quite different, as it would have quenched nuclear reactions 2~Gyr ago and already expanded into a giant star. 

\begin{figure*}
    \centering
    \includegraphics[width=0.8\linewidth]{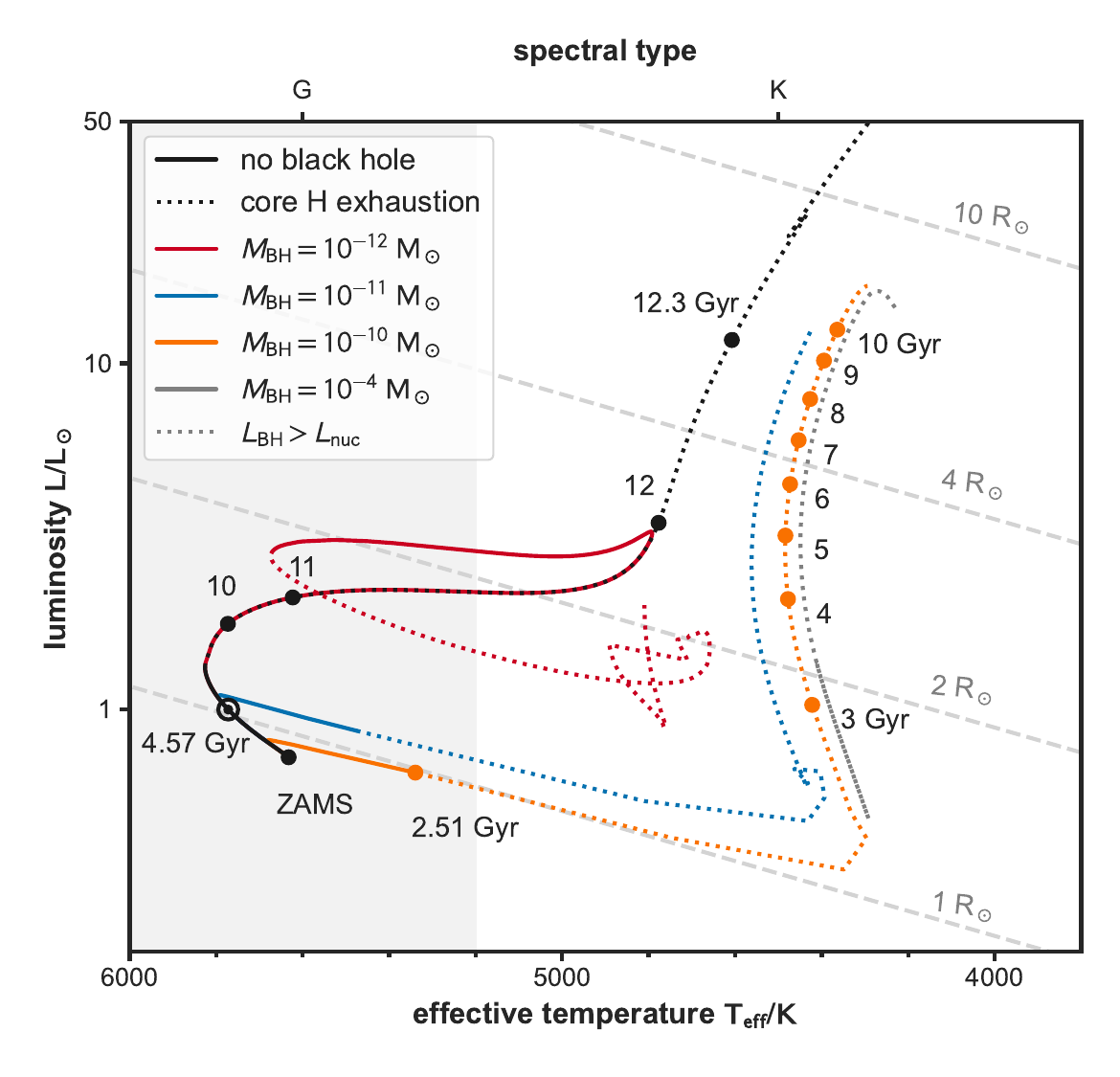}%
    \caption{Hertzsprung--Russell diagram showing the main-sequence and subsequent evolution of the Sun with and without a central black hole. 
    The solid black line representing the normal case becomes dotted when the Sun has exhausted its central hydrogen abundance. 
    The tracks with a black hole become dotted when the accretion luminosity exceeds the luminosity from fusion. 
    In all cases the star expands into a red giant. 
    Stars with lower black hole masses than those shown proceed through these evolutionary phases as normal. 
    Ages and the zero-age main sequence (ZAMS) are indicated; notice that a star with a black hole dims rapidly before climbing redward of the RGB slowly over 7~Gyr. The normal Sun climbs the RGB much quicker over a span of about 50~Myr. 
    \label{fig:hr}}
\end{figure*}

Figure~\ref{fig:hr} shows the evolution of these models in the Hertzsprung--Russell diagram. 
The unique and long-lived post main sequence phase (i.e., after fusion has ceased) may motivate observational searches.
In normal main-sequence (MS) evolution, the luminosity grows and the star migrates up the HR diagram. 
These models follow a rather different path. 
The accretion energy causes the star evolve essentially backwards along the pre-main sequence (PMS) track of standard single-star evolution, though persisting in those stages for much longer. 
As nuclear reactions begin to shut off, these stars initially migrate toward lower luminosities and cool to temperatures as low as 4300~K. 
After the complete cessation of nuclear fusion, the star then grows in luminosity and radius, while remaining up to hundreds of \mb{kelvin} cooler than the normal red giant branch. 
The temperature in this phase depends on the composition, which in turn depends on how much helium was produced during the MS evolution before entering the accretion-driven phase. 

Stars in this sparsely-populated region of the HRD are known as sub-subgiants and red stragglers. 
Hundreds of examples are now known from detailed studies of color-magnitude diagrams \citep{2017ApJ...840...66G, 2017ApJ...840...67L, 2022ApJ...927..222L} and many are known to be RS~CVn binaries. 
\citet{2017ApJ...840...67L} created evolutionary models showing three plausible explanations for these stars: binary mass transfer, envelope stripping, and magnetic activity. 
Nevertheless, they form an interesting sample in which to potentially search for stars harboring BHs.

Evolution of a star with a central BH in the red straggler phase may proceed slowly over multiple Gyr, with the timescale set by the accretion onto the BH. 
In contrast, normal red giants climb the red-giant branch proceeds over a span of less than 1~Gyr, with evolution driven by the contracting core and the presence of a hydrogen-burning shell \citep{2022ApJ...941..149M}. 
The models eventually exceed $\sim$10~$\Lsun$ and $\sim$5~$\Rsun$. 

Due to the fully convective nature of the post-MS, candidates may be identified from spectra of red giant branch stars with large He abundances. 
The Sun was about 70\% hydrogen at its birth, and will convert $\sim$10\% of its mass to helium over a period of 10~Gyr. 
If the Sun became fully mixed after the MS, its surface helium abundance would rise to about $35\%$. 
This material otherwise normally stays trapped in the core; low-mass stars furthermore also generally have a lower surface helium abundance than at their birth due to element diffusion and gravitational settling \citep[e.g.,][]{1993ApJ...403L..75C}. 

Additionally, such stars may also potentially be identified by an absence of mixed modes in their asteroseismic oscillation spectrum. 
This is because g-modes, and hence the mixed modes that are typical of red giants \citep[e.g.,][]{2017A&ARv..25....1H}, require stable stratification that is ordinarily provided by the dense radiative helium core. 
An accounting of such modes, as for instance has been done with depressed dipole modes \citep{2015Sci...350..423F}, may already be useful in constraining capture rates. 

Finally, Figure~\ref{fig:hr} also shows that for sufficiently low BH masses, the star evolves through these phases indistinguishably from a star without a BH inside it. 
For solar-mass stars with $M_{\textrm{BH}} = 10^{-12}$~\Msun{}, the star deviates from normal solar evolution only at the base of the red giant branch; for even smaller masses, the star survives the ascent up the red giant branch without significant BH growth. 
Future study will be required to understand the post-main sequence evolution of these objects, such as if they still ignite the helium flash and eventually evolve toward the white dwarf stage (see also the discussion in Section~\ref{sec:fate}). 

The models stop running when the BH reaches a mass of approximately 0.05~\Msun{}, at which point the Bondi radius is more than half the stellar radius and growing rapidly. 
The system has converted about half a percent of its total mass into radiation through the accretion process. 
\citet{2011MNRAS.414.2751B, 2012MNRAS.421.2713B} found an upper limit on the ratio of the inner BH mass to the total mass of 0.119 for supermassive quasi-stars, which they related to the Sch{\"o}nberg–Chandrasekhar limit. 
We may only speculate as to the evolution of the system beyond this point; we will return to this question in Section~\ref{sec:fate}. 

Finally, we consider a second case in which the luminosity is not limited by the Eddington limit, but rather by the maximum flux that can be carried away by convection. In this case we find that the overall evolutionary path is the same as in the first case, only vastly accelerated. Rather than the long $\sim 8$~Gyr BH-driven evolutionary phase, the star puffs into a red straggler and subsequently consumes the star over a timespan of less than 100~Myr. 
After the BH reaches a critical mass, the Sun reacts to the large injection of central energy by expanding into a giant over a thermal timescale. 
The evolution then continues over thermal timescales, which are in turn continuously shortened due to the rapidly increasing surface luminosity. 
Rather than the evolution stopping at $\sim 15$~L$_\odot$ as in the Eddington-limited case, here the star reaches $10^{4}$~L$_\odot$ before the simulation ends. 
Another difference from the Eddington-limited case is the fate of Earth: while in the previous case it was somewhat uncertain, in this case the Earth is almost certain to be engulfed, as the maximum radius of the expanding Sun grows to exceed an astronomical unit. 
As in the case of normal solar evolution, these details may change depending on the efficiency of wind-driven mass loss, which is neglected in these simulations.

\subsection{Stellar models} \label{sec:stellar} 
We have seen in the previous sections that solar models can be surprisingly long-lived for a wide range of PBH seed masses. 
Low-mass BHs have no effect on the evolution of the Sun, whereas more massive BHs bring about a unique, accretion-driven evolutionary phase. If the accretion luminosity is limited by the Eddington luminosity, then this phase may last several Gyr; otherwise, the star is consumed in tens of Myr. 

In this section, we address lifetimes more generally across a range of stellar masses. 
Because the pre-main-sequence phase is relatively short and accretion onto the BH during this phase is slow, we focus here on the main sequence. 
Similar to the analyses by \citet{2009arXiv0901.1093R} and \citet{2022MNRAS.517...28O}, we have computed a grid of zero-age main-sequence (ZAMS) models, and solved Equations~\ref{eq:edd} and \ref{eq:bondi} analytically to determine how long it takes for a BH at the center of these models to fully accrete the star. 
The details of the model grid and calculations can be found in Appendix~\ref{app:zams}. 
Here we focus on a discussion of the main findings. 

\begin{figure*}
    \centering
    \includegraphics[width=\linewidth]{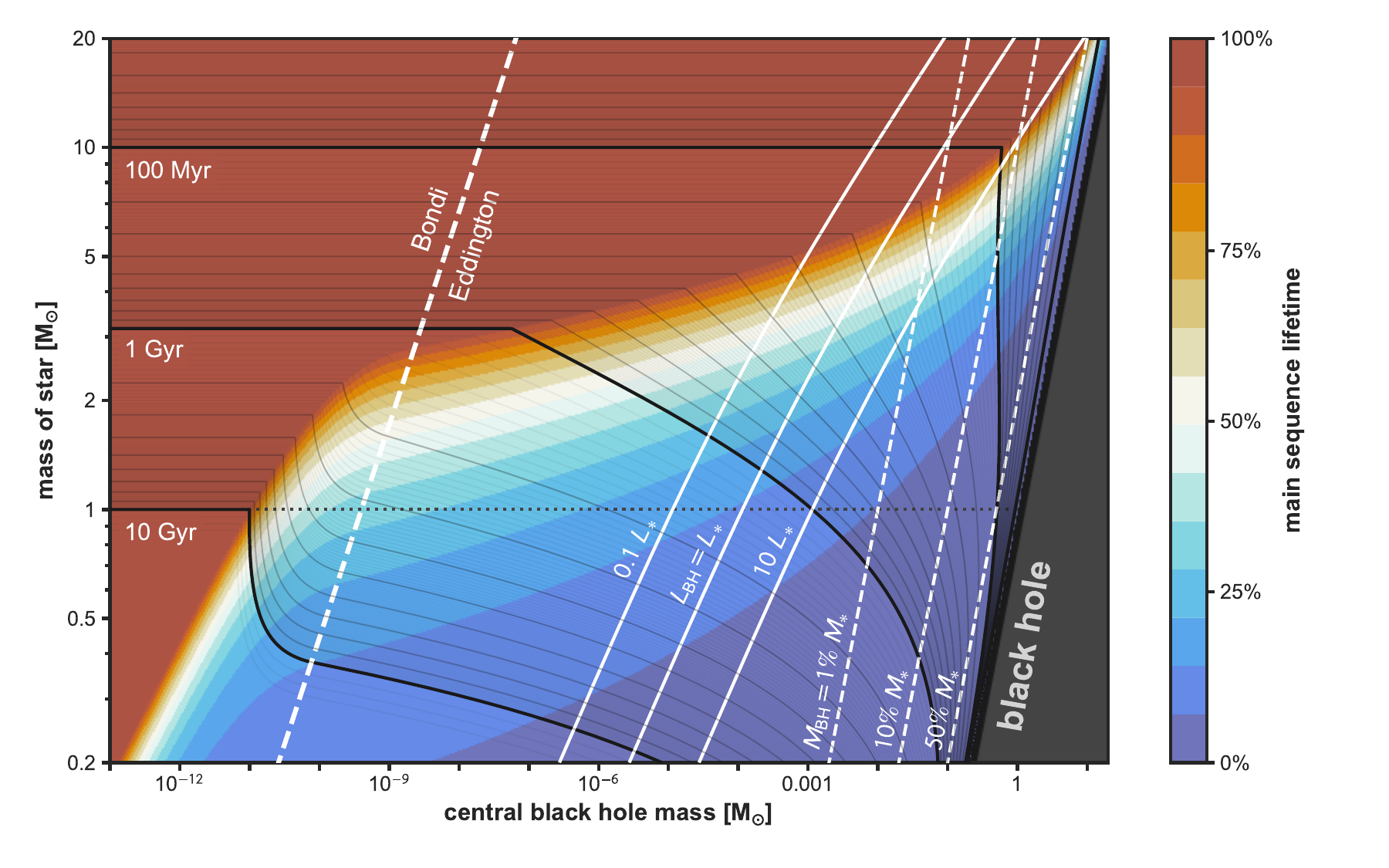}%
    \caption{Analytic accretion estimates for the lifetimes of main-sequence stars born harboring a primordial black hole. 
    Contours of constant age are given by black lines; for low-mass black holes the lifetimes of the stars are unchanged (brown shading), but as the black hole mass approaches the stellar mass, the ages go to zero (blue shading). 
    The colors indicate the percentage of the normal main-sequence lifetime that the star will live before being swallowed by the black hole; brown on the top indicates that the star will survive at least as long as its ordinary main-sequence lifetime, and white indicates that the black hole cuts the star's lifetime short by half. 
    The thick white dashed line shows the Bondi-Eddington transition, i.e., when the black hole is massive enough for its accretion luminosity to reach the Eddington limit. 
    The white lines in the middle show a comparison between the luminosity from accretion and the luminosity from fusion, showing for example that a 1~\Msun{} star could attain $1$~L$_\odot$ from slowly falling into a black hole of $10^{-4}$~\Msun{}. 
    The white dashed lines on the right show when the black hole is 1, 10, and 50\% of the stellar mass, showing for example that a 1~\Msun{} star born about a black hole of 0.1~\Msun{} could live for $\gtrsim 300$~Myr. 
    \label{fig:time}}
\end{figure*}

We must first emphasize that unlike in the previous sections, which presented detailed numerical simulations for 1~\Msun{} evolutionary tracks, these analytic estimates ignore the evolution of the internal structure of the star, as well as any interplay between the accretion luminosity and the stellar plasma. 
These approximations therefore become particularly questionable when the luminosity from the accretion exceeds the luminosity from fusion. 
Therefore, numerical simulations will be necessary to confirm these statements for additional masses other than the solar mass. 
That being said, we have found that the numerical results in Section~\ref{sec:solar} for the 1~\Msun{} case are in excellent agreement with the results found below, even for large BH masses. 
For these estimates, we assume the growth of the black hole is limited by the Eddington luminosity at a fixed radiative efficiency as in Section~\ref{sec:solar}. 
For the second case, the results would be the same at early times, but the lifetimes would be cut short around the time the Eddington limit is reached. 

Figure~\ref{fig:time} shows the lifetimes of stars harboring a central black hole. 
The mass at which the Eddington limit is reached spans from about $10^{-10}$~\Msun{} for sub-solar mass stars to $10^{-7}$~\Msun{} for 20~\Msun{} stars. 
According to the first accretion scheme, the MS lifetimes of massive stars ($M>10$~\Msun{}) would not be cut short at all unless they are born about a BH with a mass greater than $\sim$15\% of their mass, even when the BH accretion is supplying the full stellar luminosity. 
A star's longevity relative to the mass of the BH increases with increasing stellar mass. These estimates give that a $20$~\Msun{} star born about a $9$~\Msun{}~BH would live as long as the normal MS lifetime of a 20~\Msun{} star. 
Low-mass stars are especially long-lived; a $0.7$~\Msun{} star formed about a BH having $1\%$ of its mass may live $\sim$1~Gyr. 
The characteristic shape of the figure is dictated by the transition from Bondi to Eddington accretion, explaining for example why the 10~Gyr and 1~Gyr contour lines have opposite concavity. 

This figure furthermore shows us which kinds of stars are sensitive to which masses of PBHs. 
MS stars with very low mass ($0.2<M_\ast/$\Msun{}~$<3$) are potentially sensitive probes of very low-mass PBHs ($10^{-13}<$~\MBH{}$/$\Msun{}~$<10^{-9}$). 
Solar-mass stars for example would be good probes of $\sim 10^{-10}$~\Msun{} PBHs because such objects reduce the expected lifetime of the star by half if captured during formation, thus giving ample time to witness a Sun in the midst of being devoured. 
On the other hand, high-mass stars have very low ratios of density to sound speed (i.e., low erodibility), and therefore BHs in their center are unable to accrete efficiently in the Bondi regime. 
Massive MS stars ($M_\ast>10$~\Msun{}) are thus only sensitive to more massive BHs (\MBH{}~$\geq$~1~\Msun{}) over the short lifetime of their main sequence. 
However, as mentioned, these analytic estimates ignore post-MS evolution as well as the detailed interaction of the BH with the stellar plasma.
The opacity in the core drops significantly as it becomes degenerate, leading to potentially much more rapid accretion in the Eddington regime.
These detailed considerations will be the subject of future studies.

\section{Open problems} \label{sec:future}
Future work is needed to develop the theory of stellar evolution with central black holes if this method is to show promise of detecting or constraining PBHs in the asteroid-mass window. 
Here we list a few open problems that we consider to be well-posed and well-motivated. 
How many black holes were made by the Big Bang? 
How often do they become bound to star-forming clouds? 
Once captured by a cloud, how long does it take for them to reach the center of a star? 
How much energy is generated by their accretion of the stellar material, and can this luminosity be used to power the star? 
What are the final stages of stellar consumption? 
And what would be the consequences if this scenario turns out to be common? 
We will now contemplate these open problems in some detail.

\subsection{What is the initial mass function of PBHs?} 
Throughout this work, we have considered the idea of PBHs as being the solution to the dark matter problem. 
Even if they are not the full solution, they may nevertheless exist in smaller numbers, as natural scenarios appear to produce them in the early stages after the Big Bang. 

Many works that study PBHs use an idealized mass function of PBHs, in which it is assumed that all of those compact bodies are of the same mass; i.e., a so-called monochromatic mass function. 
In realistic scenarios, however, non-monochromaticity is unavoidable. 
Every known PBH production mechanism yields a spectrum of above-threshold overdensities, which then collapse to form BHs over a range of masses (see Section~II of~\citealt{Escriva:2022bwe} for an extensive discussion of various formation scenarios). 
For the most-studied case of PBH formation from inflationary overdensities, the phenomenon of critical collapse~(\citealt{Choptuik:1992jv}; see also \citealt{Koike:1995jm, Niemeyer:1997mt, Evans:1994pj, Kuhnel:2015vtw} for later work) inevitably broadens any original mass spectrum, even if the initial spectrum was monochromatic. 

After their formation, which usually happens at very early times, such as around the end of the quark epoch around $10^{-6}\,$s after the Big Bang, the PBHs also begin to accrete. 
Since these are distributed throughout the Universe, being locally inhomogeneous, the increase in mass will further broaden the initial mass and spin spectrum.
The currently most minimal and natural scenario uses critical Higgs inflation~\citep{Ezquiaga:2017fvi, Garcia-Bellido:2017mdw} in combination with the thermal history of the Universe. 
This leads to an extended PBH mass function with multiple bumps at scales of the order $10^{-6}$, $1$, $10$ and $10^{6}\,\Msun$, corresponding to the electroweak and quantum chromodynamic (QCD) scales, the pion plateau as well as to e$^+$e$^-$ annihilation, respectively, while still predicting a modest DM fraction ($\sim 10^{-3}$) in the asteroid-mass window \citep{Carr:2019kxo}. 
This mechanism has been proposed to explain the observations mentioned in the Introduction, and could potentially account for the entirety of the dark matter (see Fig.~38 of~\citealt{Carr2023Review}).

\subsection{What is the capture rate of PBHs in stars and star-forming clouds?} 
The capture rate of PBHs by stars is likely to be quite low \citep{2019JCAP...08..031M}. 
From simple kinematic arguments, a PBH in the Galactic halo falling onto a star from infinity should be accelerated above the escape velocity of that star, quickly pass through the star, and escape. 
The damping forces from accretion and dynamical friction of a small PBH are so weak that there is an effectively negligible chance that the PBH will be captured. 
Only in a neutron star are these damping forces great enough to produce an appreciable capture rate, which forms the basis of constraints on PBHs from neutron star survivability in the Galaxy \citep[e.g.,][]{2021PhRvD.103h1303B}. 

In contrast, capture during star formation is made possible by the time-dependent gravitational potential of the collapsing cloud. 
During collapse, a PBH present in the cloud is not accelerated by the gas around it, and so it can become bound to the cloud \citep{2013PhRvD..87b3507C,2014PhRvD..90h3507C}.
This capture mechanism depends on the PBH velocity distribution, with only those PBHs in the slow end of the tail being captured. 
This generally favors capture in lower mass dwarf galaxies with smaller velocity distributions over those in the Galactic disk \citep{esser2022constraints}.
One might also expect a higher rate of captures for lower PBH masses, given the greater possible number density. 
It is also possible that PBHs are clustered~(see~e.g.~\citealt{Meszaros:1975ef, 1977A&A....56..377C, 1983ApJ...268....1C, 1983ApJ...275..405F, Chisholm:2005vm, Chisholm:2011kn, Garcia-Bellido:2017xvr, Trashorras:2020mwn} or Sec.~II of \citealt{Carr2023Review} for an extensive discussion), which may allow a star to capture multiple PBHs. 
If transport toward the core is fast, it is not inconceivable that clustered PBHs may even merge in the stellar center. 
One might also expect a higher rate of captures for lower PBH masses, given the greater possible number density, making stars a powerful tool for probing the most elusive PBHs having the lowest masses. 

\subsection{Can a PBH reach the stellar core within stellar lifetimes?}
We have assumed that a PBH is already at the center by the time the star forms. 
However, unless the PBH was already located at the center during the initial collapse of a gas cloud, the PBH may have to undergo an in-spiral phase from the outer envelope to the stellar core. 
There are two main sources of drag that drive the orbit to decay: dynamical friction and hydrodynamic drag. 
If the orbiter accretes mass, this would also play a role as a drag force of roughly $\dot{M}v$, where $v$ is the speed of the orbiter. 
However, if $\dot{M}$ is the Bondi rate, the overall gravitational drag would be described by the expression of dynamical friction \citep{2011MNRAS.418.1238C, 2011MNRAS.416.3177L, 2014A&A...561A..84L}. 
Because hydrodynamic drag is proportional to the cross section of the orbiter, this contribution is subdominant in small objects like PBHs. 
Even if the effective interacting surface area is comparable to $\pi R_{\rm B}^{2}$, the hydrodynamic drag may be at most comparable to the dynamical friction. 
Thus for an order of magnitude estimate of the orbital decay time scale, we will assume dynamical friction to be the main source of decay to the orbit of the PBH. 

To take a concrete example, for a PBH with mass of $M_{\rm BH}=10^{-10}$~\Msun{} circularly plowing through an envelope layer with a local density of $\rho=10^{-4}$ g/cm$^{3}$ at a distance $r$ from the center of a $1$~\Msun{} MS star, the orbit decay time scale at that location due to dynamical friction \citep{1999ApJ...513..252O} may be estimated as:
\begin{align}
    t_{\rm decay}&\simeq 100~{\rm Myr} \left(\frac{\ln\lambda}{25}\right)^{-1}\left(\frac{M_{\rm BH}}{10^{-10}~{\rm M}_{\odot}}\right)^{-1}\nonumber\\
    \times&\left(\frac{M_{\rm en}(r)}{1~{\rm M}_{\odot}}\right)^{3/2}\left(\frac{r}{1~{\rm R}_{\odot}}\right)^{-3/2}\left(\frac{\rho(r)}{10^{-4}~{\rm g/cm}^{-3}}\right)^{-1},
\end{align}
where $M_{\rm en}$ is the enclosed mass at $r$. 
Here $\lambda$ is defined as the ratio between the effective linear size of the surrounding medium and that of the perturbing object. 
We take the former distance scale to be $r\simeq1~{\rm R}_{\odot}$ and the latter to be $R_{\rm B}$. 
With this choice, $\ln\lambda$ very weakly depends on $M_{\rm BH}$, $r$, and $M_{\rm en}$. 
For this particular case, this estimate suggests that the decay time would be short compared to the life of the star once the PBH is reasonably deep inside the envelope. 
However, it is important to remark that this estimate only gives an ``instantaneous'' decay time based on the local properties of the star. 
The total decay time would depend on a number of different factors, such as radiation from the accretion onto the perturber \citep{2017ApJ...838..103P}, equation of state \citep{2012Ap&SS.340..117K}, magnetic field \citep{2012ApJ...745..135S}, orbital motion \citep{2007ApJ...665..432K} and density gradient \citep{2001MNRAS.322...67S}.

That being said, the dependence of $t_{\rm decay}$ on $M_{\rm BH}$ and $\rho$ also suggests that low-mass PBHs, or those PBHs which could not penetrate sufficiently deep inside the envelope, could spend an exceedingly long time orbiting outside the stellar core---possibly longer than the lifetime of the star. 
This possibly long orbit decay time could have interesting implications. 
If the orbit decay time is not sufficiently shorter than the lifetime of the star, the evolution of the star would be essentially the same as that without a PBH at the core. 
For that case, the evolution of the star would in turn determine the fate of the PBH. 
The PBH's orbit decay would be significantly affected by the changes in the local density and sound speed as the star evolves. 
If the star undergoes an explosive event (such as a supernova) while the PBH orbits outside the core, the PBH could be ejected along with unbound ejecta. 
Although the wandering PBHs may not affect the evolution of the star, its non-linear orbit motion would generate a trailing pressure wave and overtake its own wake \citep{2007ApJ...665..432K}, which could affect the oscillations in the envelope. 
It is, however, difficult to speculate beyond this point; detailed hydrodynamical simulations would be a route toward making further progress on this topic.

\subsection{What is the growth rate and radiative efficiency of a microscopic black hole?} \label{sec:super}
In this work, we have considered two accretion schemes, both using a fixed radiative efficiency. 
A Bondi-like accretion rate is assumed at early times.  
In the first scheme, the luminosity is eventually limited by the Eddington luminosity; in the second, the luminosity is limited to the maximum that can be carried by convection, and thus the Bondi-like accretion rate is maintained. 
We will now discuss some of these considerations. 

Bondi-type accretion is possible when the specific angular momentum at the Bondi radius is smaller than that at the innermost stable circular orbit (ISCO). 
Assuming rigid rotation, the typical angular velocity of the Sun's interior is roughly $3\times10^{-6}$~s$^{-1}$. 
The BH mass below which the specific angular momentum at the Bondi radius ($R_{\rm BH}^{2}\Omega$, where $\Omega$ is the orbital frequency) is smaller than that at ISCO ($\sqrt{12} GM_{\rm BH}/c$) can be estimated as 
\begin{align}
    M_{\rm BH}
    \simeq 1 \Msun{}
    \left(
        \frac{c_{\rm s}}{5\times 10^7~{\rm cm~s}^{-1}}
    \right)^{4}
    \left(
        \frac{\Omega}{3\times10^{-6}~{\rm s}^{-1}}
    \right)^{-1},
\end{align}
where $c_{\rm s}$ is the sound speed at the Bondi radius. 
This justifies Bondi-type accretion at early times. 

As the flow transits toward the Eddington limit, the higher mass fluxes could slow down the outward diffusion of the radiated photons, thus trapping them and advecting along with the accretion flow. 
If the timescale for photon diffusion is larger than the free-fall timescale at the Bondi radius, then the photon trapping may suppress the radiated luminosity to below $\sim L_{\rm E}$, and the gas can continue to accrete beyond $L_{\rm E}/c^2$ \citep{1978MNRAS.184...53B}. 
As an order-of-magnitude estimate, the time for photons generated in the vicinity of the PBH to reach the Bondi radius can be estimated via the photon diffusion time $t_{\rm diff}\simeq R_{\rm B}\tau/c$, where $\tau\simeq \rho_{\rm B}R_{\rm B} \kappa$ is the optical depth and $\rho_{\rm B}$ is the density at the Bondi radius. 
This can be compared with the free-fall time $t_{\rm dyn}$ at $r\simeq R_{\rm B}$ via
\begin{align}
    \frac{t_{\rm diff}}{t_{\rm dyn}}& 
    \simeq 
    3 \left(\frac{M_{\rm BH}}{10^{-10}\Msun{}}\right)\left(\frac{\kappa}{2~{\rm cm}^{2}~{\rm g}^{-1}}\right)\nonumber\\
    &\times 
    \left(\frac{\rho_{\rm B}}{150~{\rm g ~cm}^{-3}}\right)\left(\frac{c_{\rm s, B}}{5\times10^7~{\rm cm ~s^{-1}}}\right)^{-1} 
    .
\end{align}
This order unity ratio indicates that the photon trapping may be important in determining the radiative efficiency, especially for a massive PBH embedded in a dense core. 
Therefore, limiting the mass accretion rate to Eddington rate may be unrealistic, and the BH may accrete the rest of the star much faster as in the second case that we considered. 

There are yet further difficulties with assuming the Eddington limit. 
\citet{1982MNRAS.199..833F, 1984MNRAS.206..589F} found that when gas pressure $P_g$ dominates over radiation pressure $P_r$, as is the case in the solar interior, the Eddington luminosity may lower to 
\begin{equation}
    L_{\rm F}
    =
    4\left(1-\dfrac{1}{\Gamma_1}\right)\, \dfrac{P_r}{P_g}\; L_{\rm E},
\end{equation}
which at solar conditions is approximately $0.1\%$ of the Eddington luminosity. 
If this limit holds, this implies that a BH of the same mass has much less influence on the surrounding plasma, and so the lifetime of any BH-driven phase of evolution would be considerably briefer. 
However, this reduced luminosity ignores the influence of rotation. 
Even the relatively slow rotation period of the Sun implies a nearly maximally-spinning central BH once it has accreted a significant amount of mass. 
The angular momentum of the infalling material spins up the BH, slowing the growth of the BH and raising the luminosity above $L_{\rm F}$ \citep{1995MNRAS.277...25M, 1995MNRAS.277...11M}. 
It is conceivable that a disk or torus may form, enabling even greater luminosity. 
However, the turbulent convection resulting from the accretion luminosity could prevent or disrupt the formation of such structures \citep{1995MNRAS.277...25M}. 
A significantly larger luminosity than the Flammang luminosity could also be directed along the axis of rotation, and a shear-induced instability could induce turbulent heating and further slow the accretion rate \citep{1995MNRAS.277...25M}. 
In summary, the theory of spherically symmetric accretion suggests very high accretion rates with very low accretion luminosities are possible, but the complex interplay of rotation, convection, magnetic fields, and the potential for disk or torus formation complicates the picture significantly.

If the BH does radiate at the Flammang luminosity and grow at the Bondi rate, then nearly all the infalling material solely contributes to the mass growth of the BH and almost none is converted into energy, and thus there is nearly no push back against the infalling stellar material. 
The BH still grows very slowly at early times, but upon hitting a critical mass, grows exponentially to consume the star extremely quickly. 
Since the luminosity would be low, the stellar properties would be unaffected until the BH is very massive, at which point the star has little time left before being consumed \citep{1995MNRAS.277...25M}. 
Hence in this model the Sun would not enter a long-lived accretion-powered red-giant phase if the black hole reached this critical mass on the main sequence. 

On the other hand, limiting the luminosity at the inner boundary to either the Eddington or Flammang luminosity is questionable. 
\citet{2015A&A...580A..20S} studied hydrodynamical simulations of massive stars, and found that luminosities can locally exceed the Eddington limit by a factor of a few without driving outflows, suggesting that even much larger luminosities inside the star are possible. 
In models of supermassive quasi-stars, \citet{2008MNRAS.387.1649B} take the limiting luminosity to be the Eddington limit of the entire system, rather than of the black hole alone. 
These considerations motivated our second accretion scheme, in which the luminosity is limited only by the maximum flux that can be carried away by convection. 

Lastly is the question of the radiative efficiency itself, which is highly uncertain, and depends on the properties of both the inflowing plasma and the BH itself. For instance, the radiative efficiency is expected to undergo variations on the orders of unity in relation to the spin of the BH \citep{1972ApJ...178..347B}. 
As the mass of the BH grows by orders of magnitude, if the accretion coherently adds angular momentum to the BH, then the BH spin would change rapidly, possibly approaching a limiting value, beyond which any further change may not have much effect. 

It is unclear at present how an asymmetrical flow or magnetic fields may affect the accretion. 
The differentially-rotating envelope convection zone of the Sun spins up a magnetic dynamo \citep{2002A&A...381..923S, 2020LRSP...17....4C} which complicates the situation even further. \citet{1995MNRAS.277...25M} considered how such magnetic fields may affect the formation of an accretion torus. 
Ultimately, more detailed simulations that can probe the conditions inside the Bondi sphere will be required in order to understand these complex phenomena, including the growth rates, radiative efficiencies, and super-Eddington accretion.

\subsection{How does a PBH affect the post-main sequence, and what sort of transient is associated with the final stages of stellar cannabalism by a central PBH?} \label{sec:fate}
As a star becomes a white dwarf, the core density increases along with the core electron conductivity. 
This drives the central opacities very low, which in principle would enable rapid accretion of the degenerate matter onto the BH \citep{2009arXiv0901.1093R}. 
However, this ignores the feedback from the stellar material, which may regulate this process and prevent rapid accretion. 
We plan to address later stages of stellar evolution with detailed numerical simulations in a future work. 

Our simulations end when the Bondi radius is half way to the photospheric radius, corresponding to the BH having eaten about 5\% of the stellar mass. 
The fate of the remaining material is therefore outside the domain probed by our simulations, and we may only speculate beyond this point. 
It is unclear whether the BH simply accretes the rest of the star \citep{2011MNRAS.414.2751B}. 
Instead, disk formation and outflowing jets could be possible, which would potentially carry material and angular momentum away from the BH. 
The object may also ultimately evolve toward an X-ray transient and have properties similar to the observed population of accreting X-ray binaries. 
If that is the case, then hydrodynamic simulations of the transient phase may be able to \mb{identify} characteristics that could distinguish them from those in binaries. 
Another potential observable is the disappearance of a giant star, for which upcoming transient searches such as LSST may be sensitive. 

The final black hole spin is highly uncertain and beyond the scope of the simulations presented here. 
Nevertheless, there are many interesting possibilities. 
As an order of magnitude estimate, if a PBH consumes a 1~\Msun{} star rotating with a period of 1 month, the final Kerr spin parameter ${a = J / (c\,M_{\textrm{BH}})}$ (with $J$ being the spin angular momentum) is of order 1~km, comparable to the Schwarzschild radius. 
This suggests that PBHs that consume a majority of their host star could be near maximally spinning. 
On the other hand, simulations of core-collapse supernova have shown that lower rotation rates may arise if there is significant mass loss \citep{2019ApJ...881L...1F} such as if some of the angular momentum of the infalling gas is lost through winds \citep{2020A&A...636A.104B} or jets \citep{2023arXiv230207271G}. 
Another consideration is if the PBH initially rotates rapidly. 
While most past work argues for small PBH spins at formation \citep[e.g.][]{2019JCAP...05..018D}, some calculations predict that PBHs near the classical evaporation limit may be spun up by emitting Hawking radiation, suggesting that the smallest PBHs may have large spins initially \citep{2022JHEP...12..090C}.

\subsection{What would be the consequences if this scenario turned out to be common?} \label{sec:properties}
In the relatively unlikely scenario that PBHs are both common and frequently captured by stars or star-forming regions, then they may contribute toward to several open problems. 

\citet{2023Natur.613..460L} recently found that the number of low-mass stars increases with metallicity. 
Low-mass BHs consume low-mass stars much more efficiently at low metallicity (see Figure~\ref{fig:BET}) because their cores are much less opaque, thus increasing the Eddington limit, and therefore could contribute to this effect. 

Some globular clusters show stars with helium abundances of up to 40\%, which are difficult to explain with current models \citep[e.g.,][]{2005ApJ...621L..57L}
Studies of white dwarfs in globular clusters also indicate that they have evolved from helium-rich progenitor stars \citep{2017A&A...597A..67A}. 
If early stars harboring a central BH drive a stellar wind after mixing the helium core into the envelope, then they would enrich the interstellar medium with helium for future generations of star formation. 

There appears to be a number of missing white dwarfs relative to expected numbers in some star clusters like the Hyades \citep{1974PASP...86..554T, 1975ApJ...196L.121V, 1992AJ....104.1876W, 2001MNRAS.321..199P, 2012A&A...547A..99T}. 
If accretion of degenerate plasma is much more efficient due to its very low opacity, then a large population of very low mass black holes could reveal itself in this way. 
The Hyades also appears to host several BHs \citep{2023MNRAS.524.1965T}, the presence of which may require either very small supernova kicks, or PBHs. 

The LIGO/Virgo collaborations have recently discovered numerous BH mergers, with some residing in the pair-instability mass gap \citep{2020PhRvL.125j1102A, 2020ApJ...900L..13A}. 
PBHs could in the first place naturally originate with masses within the pair-instability gap. Alternatively, these BHs could originate from PBHs of significantly lower mass that subsequently populated stars and accreted them from within. 

If PBHs constitute the dark matter, then likely \mb{PBH} production scenarios may be capable of explaining most of the LIGO/Virgo observations without a stellar origin. 
However, a sub-population of mergers \mb{with both components having $\sim 10\,\Msun$} appear to still require a different explanation \mb{than PBHs} \citep[\mb{see Figure~29 of}][]{Carr2023Review}. 
While these of course most likely originate from stellar channels, the mechanism discussed in this work could have additional bearing on this matter: by converting early stars into BHs, thereby increasing their mass, a part of the pronounced peak around a solar mass could shift towards larger masses, hence potentially yielding additional mergers in the unexplained range. 

It is possible that there is a component of the Galactic dark matter that co-orbits with the disk, a so-called \emph{dark disk}. 
Some particle theories of dark matter with weak self-interaction allow for the necessary dissipation to form a disk and would be one component of a larger dark sector \citep{2013PhRvL.110u1302F}. 
Stellar capture of PBHs can convert PBH dark matter from the halo into stellar mass BHs in the disk of the Milk Way, providing a natural formation mechanism for a Milky Way dark disk. 
Present constraints from Gaia on thin dark disks (scale heights $h\lesssim 50$~pc) suggest as much as one percent of the Galactic dark matter could be in the disk \citep{2021A&A...653A..86W,2019JCAP...04..026B}. 
A constraint on the PBH capture rate in the Milky Way could potentially be obtained from astrometric measurements.

Recent JWST observations show over-luminous objects at early times \citep{2023arXiv230401173I, 2022ApJ...937L..30L}.
Normally, a 1~\Msun{} star must wait 12~Gyr to reach 10~\Lsun{}; a BH with an initial mass of $10^{-10}$~\Msun{} could drive it to that luminosity in less than half the time. 
The ``dark star'' scenario has been evoked to explain these observations \citep{2008PhRvL.100e1101S, 2015ApJ...799..210R}, which are hypothetical stars containing WIMPs or self-interacting dark matter such as neutralino dark matter. 
On that note, we also wish to draw attention to numerous other works that have implemented candidate dark matter solutions into stellar evolution codes. 
\citet{2015JCAP...08..040V} studied asymmetric dark matter in solar models, \citet{2017PhRvD..95b3507M} extended these to other solar-type stars, and \citet{2021MNRAS.507.3434R} performed a similar study in subgiant stars. 
\citet{2016JCAP...08..062B} and \citet{2016JCAP...07..036C} studied axion cooling in pulsating white dwarfs, and \citet{2023ApJ...943...95S} carried out a similar study on the red supergiant Betelgeuse. 
\citet{2019ApJ...880L..25L} looked at the effects of WIMPS in red clump stars, and argued that a differential analysis of nearby giants and giants in the galactic center could potentially be used to distinguish stars harboring dark matter. 
\citet{2020MNRAS.491..409A} studied dark photons in RGB stars. 
More research will be required to determine which observational predictions are unique to each theory.

\section{Conclusions} \label{sec:conclusions}
Here we have numerically computed the first full solar evolution tracks with a central black hole, and also presented analytic results for a wide range of stellar masses. 
Based on our adopted accretion schemes, we find that stars harboring BHs can have surprising longevity, which broadly shows that stars are compatible with an enormous population of small BHs inhabiting the galaxy. 
Every main-sequence star living now could have formed about a PBH of less than $10^{-13}$~\Msun{} without changing the observed population. 
The present Sun could currently harbor a PBH as massive as the planet Mercury which would elude present detection capabilities. 
There is, however, no positive evidence that a BH is indeed currently in the Sun. 
If the Sun did form about a PBH, it could not have been more massive than about $10^{-11}$~\Msun{}, depending on the adopted accretion scheme and radiative efficiency, as it would otherwise change the observed properties of the present Sun. 

As stars harboring BHs are long-lived --- possibly even when the BH constitutes a significant fraction of the mass and supplies a large luminosity --- this opens a number of possible avenues for searching for the presence of such objects. 
One promising approach is through asteroseismology. 
Stars like the Sun are gardens of sound, ringing with acoustic oscillations with characteristic periods of $\sim$5 minutes. 
The frequencies of these oscillations depend intimately on both the global and interior structure of the star, and can be used for example to determine the age of the star \citep[e.g.,][]{1984srps.conf...11C, 2016ApJ...830...31B, 2020MNRAS.499.2445H}. 

When the BH mass is small, the star is essentially indistinguishable from a normal star. 
However, the accretion onto the BH causes the core of the star to become convective (see Figure~\ref{fig:kippenhahn}), mixing the material and causing the mean molecular weight gradient in that region to become flat. 
The cores of low-mass stars are ordinarily radiative, and therefore are predicted by stellar evolution theory to have steep mean molecular weight gradients. 
Asteroseismology is sensitive to the shape of this gradient \citep{2007ApJ...666..413C, 2012Natur.481...55B, 2015A&A...580A..96D, 2016A&A...589A..93D, 2018A&A...616A..24G} and has been used to characterize the masses of convective cores \citep{2020MNRAS.493.4987A}. 
Asteroseismology has provided inferences into the near-core structure of some solar-type stars \citep{2017ApJ...851...80B} including one with a convective core \citep{2019ApJ...885..143B}. 
A low-mass solar-type star with a convective core could be a signature of a central BH, as this is not produced by ordinary stellar evolution pathways, and could potentially be detectable via asteroseismology. 

A further opportunity for discovery of a star harboring a PBH becomes possible if the accretion luminosity can serve as the star's dominant source of energy. 
Here the star first becomes a sub-subgiant and later a very cool red giant, known as a red straggler. 
Ordinary red giants have a compact radiative core encased in a convective envelope and oscillate in so-called mixed modes, which is a coupling between pressure mode oscillations in the envelope and gravity mode oscillations in the core \citep[e.g.,][]{2017A&ARv..25....1H}. 
However, these latter kinds of oscillations require stably stratified regions somewhere in the star; if the star is fully convective, then no gravity modes and hence no mixed modes are possible. 
Therefore, a giant star pulsating in pure pressure modes may also be a signature of a star harboring a black hole at its center. 

On the other hand, if the accretion is radiatively inefficient, then the star would experience almost no outward change to its appearance until relatively soon before it is destroyed \citep{1995MNRAS.277...25M}. 
The star would appear as normal until at most thousands of years before its destruction, at which point its luminosity would begin to increase precipitously. 
These objects could then potentially be discovered by their rapid disappearance. 
It is also conceivable that the stellar envelope could be ejected in the end stages, leading to a sub-solar mass black hole in a nebula. 


In a future paper, we aim to perform a detailed asteroseismic characterization of stars being powered by PBHs. 
If they present a unique signature, then these objects could potentially be discovered through the data archives of the CoRoT \citep{2009A&A...506..411A}, \emph{Kepler} \citep{2010Sci...327..977B}, and TESS \citep{2015JATIS...1a4003R} missions.
Currently there is high-quality data for approximately 100 solar-type stars and thousands of giants; presently-available data may already be of sufficient quality already to find such an object. 
By the end of the decade, orders of magnitude more asteroseismic targets will become available thanks to the upcoming PLATO \citep{2014ExA....38..249R, 2017AN....338..644M} and Roman \citep{2015JKAS...48...93G} missions, including asteroseismology of $\sim$~1 million red giant stars near the galactic center. 
This presents an opportunity to either discover such objects, or to place bounds on their number and capture rate. 
The implications for stars in more advanced evolutionary stages, numerical results for stars of different masses and metallicities, and investigations into stellar populations will also be explored in future works. 

\vspace{\baselineskip}

There is a question of what to call these hypothetical low-mass quasi-stars, should they ever be found. They might be called ``Hawking~stars.''

\acknowledgements{\noindent The authors thank Manolis Manussakis for hospitality. 
We are grateful to Andrei Beloborodov, Stephen Justham, Aniket Bhagwat, Christian Partmann, Carles Badenes, Alison Sills, Mitchell Begelman, and the anonymous referee for useful discussions. 
Funding for the Stellar Astrophysics Centre was provided by The Danish National Research Foundation (Grant DNRF106). Financial support for this publication comes from Cottrell Scholar Award \#CS-CSA-2023-139 sponsored by Research Corporation for Science Advancement.
} 

\appendix 
\section{ZAMS models} \label{app:zams}
\begin{figure}
    \includegraphics[width=0.33\textwidth]{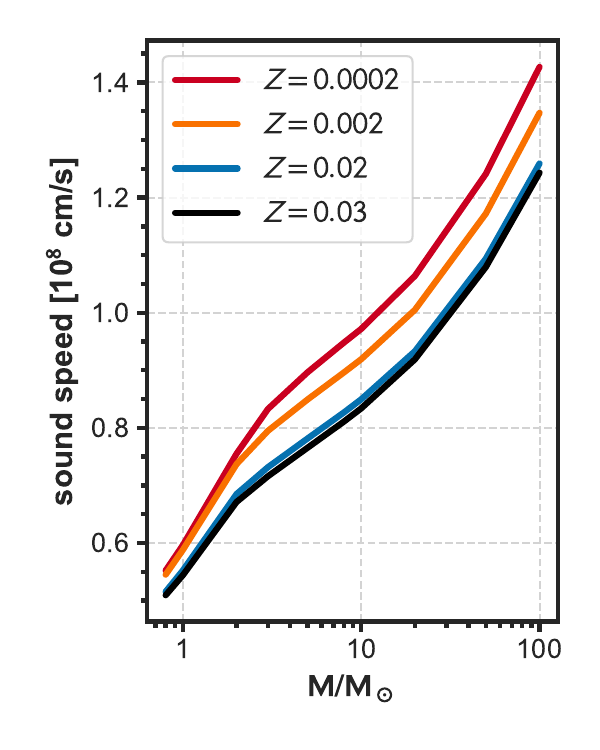}%
    \includegraphics[width=0.33\textwidth]{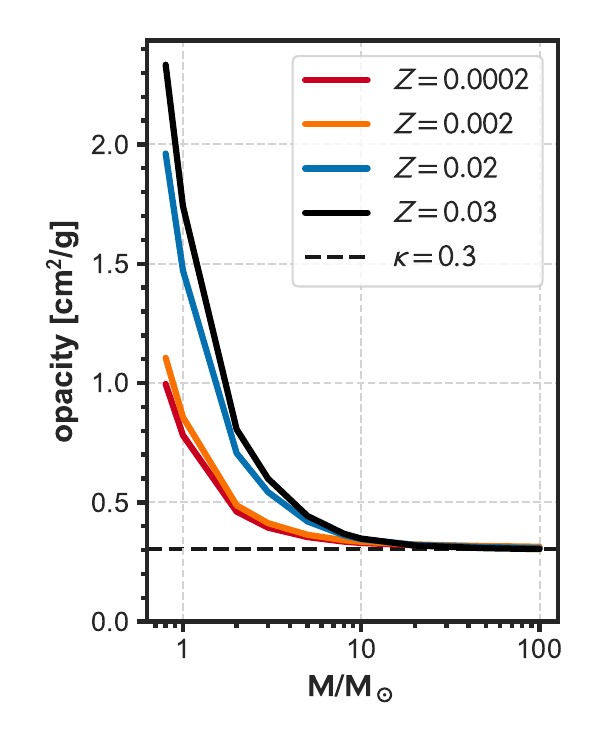}%
    \includegraphics[width=0.33\textwidth]{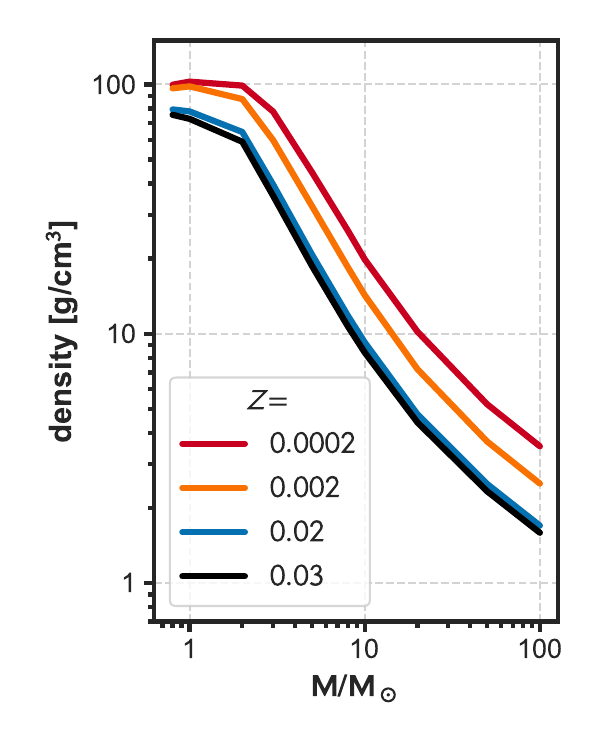}%
    \caption{Central sound speed (left), opacity (center), and density (right) for ordinary zero-age main sequence models as a function of mass and shown for $1\%$, $10\%$, $100$\%, and $150$\% of the solar metallicity. \label{fig:zams}}
\end{figure}

To facilitate accretion timescale estimates, we show the central sound speed and central opacity of our models at ZAMS in Figure~\ref{fig:zams}. 
Bondi accretion is inversely related with sound speed and Eddington accretion is inversely related with opacity. 
The central sound speed only changes by a factor $\sim$2 with increasing stellar mass around a typical value of $10^8$~cm$/$s.
The opacity decreases with increasing stellar mass, eventually asymptoting for $M>10~\Msun$ at a minimum of $0.3$~cm$^2/$g. 
For comparison, past authors have assumed fully ionized H using $\kappa= \sigma_T / m_p = 0.4$~cm$^2/$g, with Thompson cross section $\sigma_T$ and proton mass $m_p$. 
Central densities vary by about two orders of magnitude across stellar masses. 

We find these central values are reasonably well fit by: 
\begin{align}
    c_s &= 1.6 \times 10^7\, (3.5 + \log_{10}M_\ast)
    \, ,\\
    \kappa &= 0.3 + 1.2\times M_\ast^{-1.3}
    \, ,\\
    \log_{10} \rho &= \dfrac{-1.9}{1 + \exp\{- 2.7 (\log_{10} (M_\ast) - 1) \}} + 2
    \, ,
\end{align}
where $M_\ast$ is in \Msun{} units and the rest are in cgs. 
Metallicity has at most a 15\% effect on sound speed, whereas a high metallicity can more than double the opacity in the core of a low-mass star. Note that we have neglected the Z dependency in developing these formulas. 

\begin{figure}
    \includegraphics[width=0.33\textwidth]{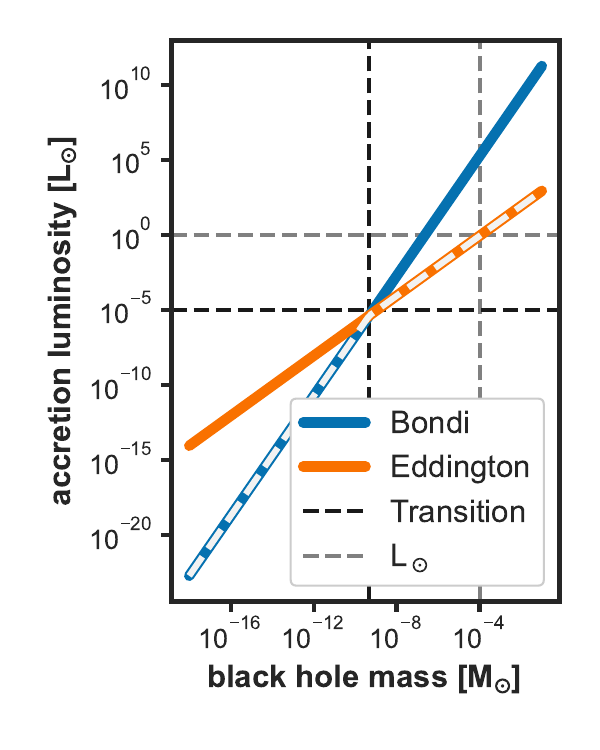}%
    \includegraphics[width=0.33\textwidth]{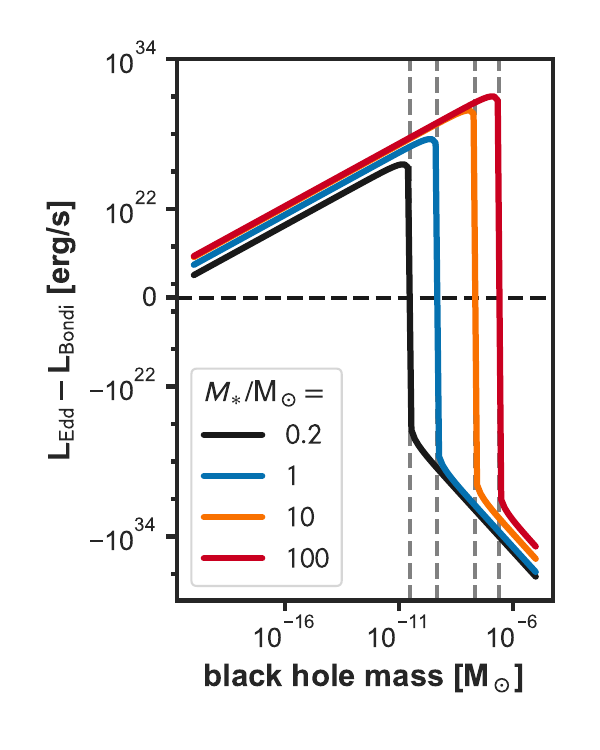}%
    \includegraphics[width=0.33\textwidth]{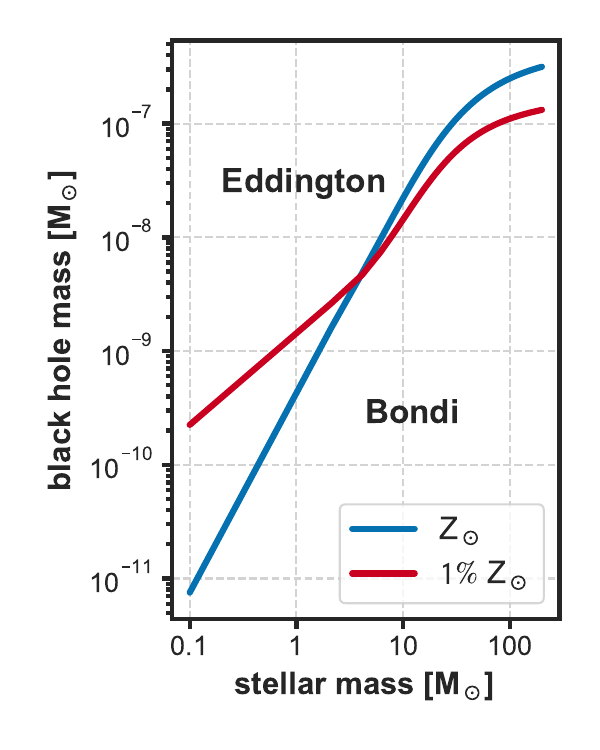}%
    \caption{ The transition between Bondi and Eddington accretion onto the central BH for zero-age main-sequence stars. 
    Left panel: the accretion luminosity of a $1$~\Msun{} star at ZAMS (white dashed line) as a function of the mass of the PBH that it formed about. 
    Middle panel: the difference in Bondi and Eddington luminosities at ZAMS. 
    Right panel: the transition mass between Bondi and Eddington accretion at ZAMS. 
    For small BHs, the accretion proceeds at the Bondi rate (Equation~\ref{eq:bondi}); larger BHs accrete at the Eddington limit (Equation~\ref{eq:edd}).  More massive stars have significantly higher Bondi-Eddington transition masses. Metallicity is very important in determining this transition in low-mass stars. 
    \label{fig:BET}}
\end{figure}

We use these formulas to estimate the transition point between Eddington and Bondi accretion for a ZAMS star by equating Equation~(\ref{eq:edd}) and Equation~(\ref{eq:bondi}). The results are shown in Figure~\ref{fig:BET}. This relation at solar metallicity is well-approximated by 
\begin{equation}
    M_{\textrm{BET}} = 4\times 10^{-10} M_\ast^{1.7}
    \, .
\end{equation}
All stars that form about a PBH whose mass at capture is at least 1-2 orders of magnitude lower than the $M_{\textrm{BET}}$ will survive at least the main sequence; and stars above 3~\Msun{} survive even if the BH mass is significantly greater than the $M_{\textrm{BET}}$. 
We aim to assess its post-main sequence fate in a future work, including the interesting feedback and self-regulation processes arising between the interplay of the accretion onto the black hole and the response of the stellar plasma. 
Note that the densities of stellar cores increase substantially throughout MS evolution; for instance, the solar core had a ZAMS density of $\sim80$~g$/$cm$^3$, whereas now it has a density of $\sim150$~g$/$cm$^3$.
We ignore these effects in this analytical estimate, but they are incorporated in the numerical one presented in the main text. 

\section{Analytic timescale} \label{app:anal}
We are now prepared to estimate the time it takes for a PBH to fully accrete the star, and compare this to the MS age. 
To reasonable accuracy, MS ages follow
\begin{equation} \label{eq:t_ms}
    t_{\textrm{MS}} = 10~M_\ast^{-2}~\textrm{Gyr}
    \, .
\end{equation}
For the Bondi case, we will estimate the time it takes $t_\textrm{Bondi}$ for the BH mass to grow until it reaches the Eddington limit via
\begin{equation}
    \int_{M_{\textrm{BH},0}}^{M_{\textrm{BET}}}
        \dfrac{1}{M_{\textrm{BH}}^2}
    \;\textrm{d} M_{\textrm{BH}}
    =
    -\left(
        \dfrac{1}{M_\textrm{BET}}
        -
        \dfrac{1}{M_{\textrm{BH},0}}
    \right)
    =
    \dfrac{t_{\textrm{Bondi}}}{S_{\textrm{Bondi}}}
    \, ,
\end{equation}
where $M_{\textrm{BH},0}$ is the BH seed mass, $M_{\textrm{BET}}$ is the mass of the Bondi-Eddington transition, and the Bondi accretion mass sensitivity is given by
\begin{equation}
    S_{\textrm{Bondi}}
    =  
    \dfrac{\epsilon}{(1-\epsilon)\eta}
    \dfrac{c^2 \Gamma_1 c_s}{16\pi G^2 \rho}
    \, .
\end{equation}
For the Eddington case, we have
\begin{equation}
    \int_{M_m}^{M_\ast}
        \dfrac{1}{M_{\textrm{BH}}}
    \;\textrm{d} M_{\textrm{BH}}
    =
    \dfrac{t_{\textrm{Edd}}}{S_{\textrm{Edd}}}
    \, ,
\end{equation}
where $M_m = {\max(M_{\textrm{BH},0}}, M_{\textrm{BET}})$ and the Eddington accretion mass sensitivity is
\begin{equation}
    S_{\textrm{Edd}}
    = 
    \dfrac{\epsilon}{1-\epsilon}
    \dfrac{c \kappa}{4\pi G}
    \, ,
\end{equation}
which implies that 
\begin{equation}
    t_{\textrm{Edd}} 
    = 
    S_{\textrm{Edd}} \log \left(
        \dfrac{M_\ast}{M_{m}}
    \right)
    .
\end{equation}
Finally, we used these analytic estimates to construct Figure~\ref{fig:time}. 
The characteristic shape of the curve in this figure (i.e., the point at which the curve begins to transition from red to blue) is given by equating the MS lifetime $t_{\textrm{MS}}$ with the total accretion lifetime.

\bibliographystyle{aasjournal.bst}
\bibliography{main}

\end{document}